\documentclass[ba, preprint]{imsart}

\RequirePackage{amsthm,amsmath,amsfonts,amssymb}
\RequirePackage[authoryear]{natbib}
\RequirePackage[colorlinks,citecolor=blue,urlcolor=blue,backref=page,backref=page]{hyperref}
\RequirePackage{graphicx}
\usepackage{bm}
\usepackage{pdfpages}

\pubyear{2024}
\volume{TBA}
\issue{TBA}
\firstpage{1}
\lastpage{1}

\startlocaldefs
\theoremstyle{plain}

\theoremstyle{definition}

\theoremstyle{remark}

\newcommand{\dat}{\mathcal{D}}
\newcommand{\x}{\bm{x}}
\newcommand{\y}{\bm{y}}

\newcommand{\fbold}{\bm{f}}
\newcommand{\fboldj}[1]{\fbold^{(#1)}}

\newcommand{\K}{\bm{K}}

\newcommand{\Kj}[1]{\K^{(#1)}}

\newcommand{\secname}{Sec.}
\newcommand{\ename}{Eq.}
\newcommand{\fname}{Fig.}

\newcommand{\param}{\bm{\theta}}
\newcommand{\paramGP}{\param_{\text{GP}}}
\newcommand{\paramOBS}{\param_{\text{obs}}}

\endlocaldefs

\begin{document}

\begin{frontmatter}
\title{Scalable mixed-domain Gaussian process modeling and model reduction for longitudinal data}
\runtitle{Scalable mixed-domain GPs}

\begin{aug}
\author[A]{\fnms{Juho}~\snm{Timonen}\ead[label=e1]{juho.timonen@aalto.fi}\orcid{0000-0003-2341-6765}} \and
\author[A]{\fnms{Harri}~\snm{Lähdesmäki}\ead[label=e2]{harri.lahdesmaki@aalto.fi}}

\address[A]{Department of Computer Science,
Aalto University\printead[presep={,\ }]{e1,e2}}

\runauthor{Timonen \& Lähdesmäki}
\end{aug}

\begin{abstract}
Gaussian process (GP) models that combine both categorical and continuous input variables have found use in analysis of longitudinal data and computer experiments. However, standard inference for these models has the typical cubic scaling, and common scalable approximation schemes for GPs cannot be applied since the covariance function is non-continuous. In this work, we derive a basis function approximation scheme for mixed-domain covariance functions, which scales linearly with respect to the number of observations and total number of basis functions. The proposed approach is naturally applicable to also
Bayesian GP regression with discrete observation models. We demonstrate the scalability of the approach and compare model reduction techniques for additive GP models in a longitudinal data context. We confirm that we can approximate the exact GP model accurately in a fraction of the runtime compared to fitting the corresponding exact model. In addition, we demonstrate a scalable model reduction workflow for obtaining smaller and more interpretable models when dealing with a large number of candidate predictors.
\end{abstract}



\end{frontmatter}

\section{Introduction}

Gaussian processes (GPs) offer a flexible nonparametric way of modeling unknown functions. While Gaussian process regression and classification are commonly used in problems where the domain of the unknown function is continuous, recent work has seen use of GP models also in mixed domains, where some of the input variables are categorical or discrete and some are continuous. Applications of mixed-domain GPs are found e.g.\ in Bayesian optimization \citep{garridomerchan2018}, computer experiments \citep{qian2008, deng2017, roustant2018, wang2021, zhang2015} and longitudinal data analysis \citep{cheng2019, timonen2021}. For example in biomedical applications, the modeled function often depends on categorical covariates, such as treatment vs.\ no treatment, and accounting for such time-varying effects is essential.  Non-trivial mixed-domain kernels are required in order to obtain an interpretable model with both shared and category-specific effects \citep{timonen2021}.

Since all commonly used kernel functions (i.e.\ covariance functions) are defined for either purely continuous or purely categorical input variables, kernels for mixed-domain GPs are typically obtained by combining continuous and categorical kernels through multiplication. Additional modeling flexibility can be obtained by summing the product kernels as has been done in the context of GP modeling for longitudinal data \citep{cheng2019, timonen2021}. 

It is well known that exact GP regression has a theoretical complexity of $\mathcal{O}(N^3)$
and requires $\mathcal{O}(N^2)$ memory, where $N$ is the number of observations. This
poses a computational problem which in practice renders applications of exact GP regression infeasible for large data. Various scalable approximation approaches for
GPs have been proposed (see e.g.\ \citep{liu2020} for a review). However, many popular approaches, such as the inducing point \citep{snelson2006, titsias2009} and
kernel interpolation \citep{wilson15} methods, can only be applied directly if the kernel (i.e.\ covariance) function is continuous and differentiable. In addition, they typically require the Gaussian observation model, which is not appropriate for modeling 
for example discrete, categorical or ordinal response variables. See \secname~\ref{s: relatedwork} for a review of previous methods.

In the presence of a variety of possible explanatory covariates, reducing the model to a minimal set of variables can be an essential part of building an interpretable and a useful model. The combinatorial explosion of possible alternative
models as a function of the number of candidate covariates poses an additional problem for scalable modeling workflow that aims to produce a small an interpretable model with good predictive properties.

In this work, we present a scalable approximation and model reduction scheme for additive mixed-domain GPs, where the covariance structure depends on both continuous and categorical variables. We extend the Hilbert space reduced-rank approximation \citep{solin2019} for said additive mixed-domain GPs, making it applicable to e.g.\ analysis of large longitudinal data sets. The approach scales linearly with respect to data set size and allows a wide variety of categorical kernels that specify possible correlation over groups. It allows an arbitrary observation model and full Bayesian inference for the model hyperparameters, and is suitable for longitudinal data as it allows product kernels of continuous and categorical kernels for modeling for example group-specific effects that sum to zero. Furthermore, we demonstrate how to use the projection predictive technique (see e.q.\ \citep{pavone2020}) for said models and compare it with a variance decomposition based covariate relevance assessment technique \citep{timonen2021} to obtain recommendations how to efficiently produce small and interpretable additive GP models for longitudinal data. We offer an open source R package \citep{rcore2023} called \texttt{lgpr2}\footnote{\url{https://github.com/jtimonen/lgpr2}} that implements the described longitudinal approximate GP models and model reduction techniques.

\section{Background}
\label{s: gp}

\subsection{Gaussian processes}
A Gaussian process (GP) is a collection of random variables, any finite number of which has a multivariate normal distribution \citep{rasmussen2006}. A function $f$ is a GP
\begin{equation}
f \sim \mathcal{GP}\left(m(\bm{x}), k(\bm{x}, \bm{x}') \right)
\end{equation}
 with mean function $m(\x)$ and kernel (or covariance) function $k(\x, \x')$, if for any finite number of inputs $\{\x_i \in \mathcal{X} \}_{i=1}^N$, the vector of function values $\fbold = \left[f(\x_1), \ldots, f(\x_N) \right]^\top$ follows a multivariate normal distribution $\fbold \sim \mathcal{N}\left(\bm{m},\K \right)$ with mean vector $\bm{m} = \left[m(\x_1), \ldots, m(\x_N) \right]^\top$ and $N \times N$ covariance matrix $\K$ with entries $\{ \K \}_{nm} = k(\x_n, \x_m)$. The mean function is commonly the constant zero function $m(\x) = 0$, and we have this convention throughout the paper. The kernel function encodes information about the covariance of function values at different points,
 and therefore affects the model properties crucially.

\subsection{Bayesian GP regression}
\label{s: bayesian_gp}
In GP regression, the conditional distribution of response variable $y$ given covariates $\x$ is modeled as some parametric distribution $p(y \mid f(\x), \paramOBS)$, where $\paramOBS$ represents possible parameters of the observation model. The function $f$ has a zero-mean GP prior with covariance function $k(\bm{x}, \bm{x}' \mid \paramGP)$ that has hyperparameters $\paramGP$. We focus on Bayesian GP modeling, where in addition to the GP prior for $f$, we have a parameter prior distribution $p(\param)$ for $\param = \left\{ \paramGP , \paramOBS \right\}$. Given $N$ observations $\dat = \{y_n, \x_n\}_{n=1}^N$ our goal is to infer the posterior 
\begin{equation}
\label{eq: joint_posterior}
p\left(\param, \fbold \mid \mathcal{D}\right) \propto p\left(\param, \fbold\right) \cdot p(\y \mid \fbold, \param),
\end{equation}
where $\y = \left[y_1, \ldots, y_N\right]^{\top}$. The part
\begin{equation}
\label{eq: joint_prior}
     p\left(\param, \fbold\right) = p\left(\fbold \mid \param\right) \cdot p(\param)
\end{equation}
is the prior and
\begin{equation}
\label{eq: likelihood}
     p(\y \mid \fbold, \param) = \prod_{n=1}^N p(y_n \mid f(\x_n), \paramOBS)
 \end{equation}
is the likelihood. This task often has to be done by sampling from $p\left(\param, \fbold \mid \mathcal{D}\right)$ using MCMC methods, which requires evaluating the right-hand side of \ename~\ref{eq: joint_posterior} (and possibly its gradient) thousands of times. As the likelihood and parameter prior usually are independent over each parameter and data point, they scale linearly and are not a bottleneck. Instead, computing the GP prior density
\begin{equation}
\label{eq: f_prior}
p\left(\fbold \mid \param\right) = \mathcal{N}\left(\fbold \mid \bm{0}, \K \right),
 \end{equation}
 where the $N \times N$ matrix $\K$ has entries $\{ \K \}_{nm} = k(\x_n, \x_m \mid \paramGP)$, is a costly operation as evaluating the (log) density of the $N$-dimensional multivariate normal distribution has generally $\mathcal{O}(N^3)$ complexity (see \secname~S3.6). Furthermore, the matrix $\K$ takes $\mathcal{O}(N^2)$ memory.
 
 An often exploited fact is that if the observation model (and therefore
 likelihood) is Gaussian, $\fbold$ can be analytically marginalized and only the marginal posterior $p\left(\param \mid \mathcal{D}\right)$ needs to be sampled. This reduces the MCMC dimension by $N$ and likely improves sampling efficiency, but one however needs to evaluate $p(\y \mid \param)$ which is again an $N$-dimensional multivariate Gaussian. The $\mathcal{O}(N^3)$ complexity and $\mathcal{O}(N^2)$ memory requirements therefore remain. In this paper, we generally assume an arbitary observation model, and defer the details of the Gaussian observation model until \secname~S3.2.

\subsection{Additive GP regression}

In additive GP regression, the modeled function $f$ consists of additive components so
that $f = f^{(1)} + \ldots + f^{(J)}$, and each component $j = 1, \ldots, J$ has a GP prior
\begin{equation}
\label{eq: gp}
f^{(j)} \sim \mathcal{GP}\left(0, k_j(\x, \x') \right),
\end{equation}
independently from other components. This means that the total GP prior is 
$f \sim \mathcal{GP}\left(0, k(\x, \x')  \right)$ with
\begin{equation}
\label{eq: additive_kernel}
    k(\x, \x') = \sum_{j=1}^J k_j(\x, \x’).
\end{equation}
Furthermore, $\fboldj{j} \sim \mathcal{N}\left(\bm{0}, \Kj{j} \right)$ for each vector
\begin{equation}
    \fboldj{j} = \left[f^{(j)}(\x_1), \ldots, f^{(j)}(\x_N) \right]^\top,
\end{equation}
where $j = 1, \ldots, J$. The matrix $\Kj{j}$ is defined so that its elements are $\{ \Kj{j} \}_{nm} = k_j(\x_n,\x_m)$. This means that the prior for $\fbold = \fboldj{1} + \ldots + \fboldj{J}$ is $\fbold \sim \mathcal{N}\left(\bm{0},\K \right)$, where $\K = \sum_{j=1}^J \Kj{j}$. 

In additive GP models, we are often ultimately interested in inferring the components $\fboldj{j}$, in which case Bayesian inference requires sampling all $\fboldj{j}$. Adding one component increases the number of parameters by $N$ (plus the possible
additional kernel hyperparameters). Moreover, the 
multivariate normal prior (\ename~\ref{eq: f_prior}) needs to be evaluated for each component,
adding the computational burden. In the case of Gaussian likelihood, adding more components does not add any multivariate normal density
evaluations as it still needs to be done only for $p(\y \mid \param)$. Also the marginal posteriors of each $\fboldj{j}$ are analytically available (see  \secname~S3.2).

\subsection{Mixed-domain kernels for longitudinal data}
\label{s: mixed_domain_gps}
Longitudinal data is common in biomedical, psychological, social and other studies and consists of multiple measurements of several subjects at multiple time points. In addition to time (often expressed as subject age), other continuous covariates can
be measured. Moreover, in addition to subject id, other categorical covariates, such as
treatment, sex or country can be available. In the statistical methods literature, such data is commonly modeled using generalized linear mixed effect models \citep{verbeke2000}. In recent work \citep{cheng2019, quintana2016, timonen2021}, longitudinal data has been modeled using additive GPs, where, similar to commonly used linear models, each component $f^{(j)}$ is a function  of at most one categorical and one continuous variable. Each variable is assigned a one-dimensional base kernel and for components that contain both a continuous and categorical kernel, the kernel $k_j$ is their product. As the total kernel $k$ is composed
of the simpler categorical and continuous kernels through multiplication and addition,
it has a mixed domain.

These models have the very beneficial property that the effects of individual covariates
are interpretable. The marginal posterior distributions
of each component can be studied to infer the marginal effect of different covariates. As an example, if $k_j$ is just the exponentiated quadratic (EQ) kernel
\begin{equation}
    \label{eq: eq_kernel}
    k_{\text{EQ}}(x, x') = \alpha^2 \exp \left(-\frac{(x-x')^2}{2\ell^2} \right)
\end{equation}
and $x$ is age, the component $f_j$ can be interpreted as the shared effect of age. 
On the other hand, if $k_j$ is the product kernel  $k_{\text{EQ}\times\text{ZS}}{\left((x,z), (x',z')\right)} = k_{\text{EQ}}(x, x')k_{\text{ZS}}(z, z')$
where
\begin{equation}
    \label{eq: zs_kernel}
    k_{\text{ZS}}(z, z') = 
    \begin{cases}
       1 \hspace{0.77cm} \text{ if } z = z' \\
       -\frac{1}{C-1} \text{ if } z \neq z'
    \end{cases}
\end{equation}
is the zero-sum (ZS) kernel \citep{kaufman2010} for a categorical variable
$z$ that has $C>1$ categories, $f^{(j)}$ can be interpreted as the category-specific effect of the continuous covariate $x$. This also has the property that the effect sums to zero over categories at all values for $x$ (see \cite{timonen2021} for proof), which helps in separating the category effect from the shared effect, if a model has both. 


\section{Related research}
\label{s: relatedwork}

\paragraph{GPs and categorical inputs}
\label{s: related_categorical}
A suggested approach to handle GPs with categorical covariates is to use a one-hot encoding which turns a variable with $C$ categories into $C$ binary variables, of which only one is on at a time, and then apply a continuous kernel for them. \cite{garridomerchan2018} highlight that the resulting covariance structure
is problematic because it does not take into account that only one of the binary variables can be one at a time. This poorly motivated approach might have originated merely from the fact that common GP software have lacked support for categorical kernels. We find it more sensible to define kernels directly on categorical covariates, as that way we can always impose the desired covariance structure.

Category-specific effects of a continuous covariate $x$ can be achieved also by assigning independent GPs for the different categories. This way we have only continuous kernel functions, and can possibly use scalable approaches that are designed for them. This limited approach however cannot define any additional covariance structure between the categories, such as the zero-sum constraint (\ename~\ref{eq: zs_kernel}). The ZS kernel is a special case of compound
symmetry (CS), and for example \cite{roustant2018} concluded that
a CS covariance structure was more justified than using only indenpendent GPs in their nuclear engineering application. 

\cite{chung2020} developed a deep mixed-effect GP model that facilitates individual-specific
effects and scales as $\mathcal{O}(PT^3)$, where $P$ is the number of individuals and $T$ is the number of time points. \cite{zhang2020} handled categorical inputs by mapping them
to a continuous latent space and then using a continuous kernel. While this approach
can detect interesting covariance structures, it does not remove need to perform
statistical modeling with a predefined covariance structure as in
\secname~\ref{s: mixed_domain_gps}. Another related non-parametric way to model group effects is to use hierarchical generalized additive models \citep{pedersen2019}, as smoothing splines can be seen as a special case of GP regression \citep{kimeldorf1970}.

\paragraph{Computer experiments}
In analysis of computer experiments \citep{sacks1989}, GPs can be used as a surrogate model for the output of a computer code that takes a long time to run. Often, both qualitative and quantitative factors affect the output of a computer code. Whereas the kernels used in this paper are valid by construct, a strategy to find general valid
covariance functions for computer experiments involving both continuous and categorical variables
was developed by \cite{qian2008}, using semidefinite programming. The output of the computer code can be assumed to be observed exactly, meaning that there is no model for the observation noise (\secname~\ref{s: bayesian_gp}). In this case,
the presented approximation methodology can still be used via the strategy detailed in \secname~S3.3. with $\sigma^2 = 0$, as no noise can be seen as a special case of the Gaussian observation model with $\sigma^2 \rightarrow 0$.

\paragraph{Scalable GP approximations}
\label{s: related_scalable}
A number of approximation methods exist that reduce the complexity of GP regression to
$\mathcal{O}(N \cdot M^2)$, where $M$ controls the accuracy of the approximation. Popular approaches rely on global sparse approximations \citep{quinonerocandela2005}
of the $N \times N$ covariance matrix between all pairs of data points, using $M$ inducing points. The locations of these inducing points are generally optimized using gradient-based continuous optimization simultaneously with model hyperparameters,
which cannot be applied when the domain is not continuous. In \cite{fortuin2021}, the inducing-point approach was studied in purely discrete domains and \cite{cao2015}
presented an optimization algorithm that alternates between discrete optimization of inducing points, and continuous optimization of the hyperparameters. Disadvantages of this method are that it cannot find inducing points outside of the training data, does not perform full Bayesian inference for the hyperparameters, and assumes a Gaussian observation model.

A Hilbert space basis function approach for reduced-rank GP approximation in continuous domains, on which this work is based, was proposed by \cite{solin2019}. Its use in the practice in the Bayesian setting was studied more in \cite{riutortmayol2020}. 

\section{Mixed-domain covariance function approximation}
\label{s: our_approach}

\subsection{Basic idea}
\label{s: basic_idea}
We continue with the notation established in \secname~\ref{s: gp}, and note that $\bm{x}$ denotes a general input that can consist of both continuous and categorical dimensions. We consider approximations $\fbold \approx \tilde{\fbold}$ that decompose the GP kernel function as
\begin{equation}
\label{eq: approximation_idea}
k(\x, \x') \approx \tilde{k}(\x, \x') = \sum_{m=1}^M \psi_m(\x) \psi_m(\x'),
\end{equation}
where functions $\psi_m$ have to be designed so that the approximation is accurate but easy to compute. This is useful in GP regression, because we get a low-rank approximate decomposition for the kernel matrix $\K \approx \tilde{\K} = \bm{\Psi} \bm{\Psi}^{\top}$, where $\bm{\Psi}$ is the $N \times M$ matrix with elements $\left[ \bm{\Psi} \right]_{n,m} = \psi_m(\x_n)$.
Using this approximation, we can write the approximate GP prior $\tilde{\fbold} \sim \mathcal{N}\left(\bm{0}, \bm{\Psi} \bm{\Psi}^{\top} \right)$ using $M$ parameters $\xi_m$ with independent standard normal priors, connected to $\fbold$ through the reparametrization $\tilde{\fbold} = \bm{\Psi} \bm{\xi}$, where $\bm{\xi} = \left[\xi_1, \ldots, \xi_M\right]^{\top}$. Evaluating the prior density $p(\bm{\xi}) = \prod_{m=1}^M \mathcal{N}\left(\xi_m \mid 0, 1 \right)$ has now only $\mathcal{O}(M)$ cost. After obtaining posterior draws of $\bm{\xi}$, we can obtain posterior draws of $\tilde{\fbold}$ with $\mathcal{O}(NM)$ cost, which comes from computing the matrix $\bm{\Psi}$. The likelihood (\ename~\ref{eq: likelihood}) can then be evaluated one data point at a time and the total complexity of the approach is only $\mathcal{O}(NM + M)$. Furthermore, the memory requirement is reduced from $\mathcal{O}(N^2)$ to $\mathcal{O}(NM)$, since we only need to store $\bm{\Psi}$ and never compute $\K$. This is the approach used throughout this paper, and the focus is on how to design the functions $\psi_m$ for different kernel functions so that the approximation is accurate with $M \ll N$.
 
\subsection{Continuous isotropic covariance functions}

A continuous stationary covariance function $k(\x, \x'): \mathbb{R}^D \times \mathbb{R}^D \rightarrow \mathbb{R}$ depends only on the difference $\bm{r}=\x-\x'$ and can therefore be written as $k(\bm{r})$. Such covariance functions can be approximated by methods that utilize the spectral density
\begin{align}
\label{eq: spectral_dens}
    S(\omega) &= \int k(\bm{r}) e^{- i \omega^\top \bm{r}} \text{d} \bm{r}.
\end{align}
If the covariance function is isotropic, meaning that it depends only on the Euclidean norm $\lVert \bm{r} \rVert$, also $S(\omega)$ is isotropic and can be written as $S\left(\lVert \omega \rVert \right)$, i.e. as a function of one variable. As shown in \citep{solin2019}, an isotropic covariance function can be approximated as
\begin{equation}
    \label{eq: k_approx_trunc}
    \tilde{k}(\x, \x') = \sum_{b=1}^{B} s_b \phi_b(\x) \phi_b(\x'),
\end{equation}
where $s_b = S(\sqrt{\lambda_b})$ and $\phi_b$ and $\lambda_b$ are the $B$ first eigenfunctions and eigenvalues of the Dirichlet boundary value problem
\begin{equation}
\label{eq: dirichlet}
    \begin{cases}
      \frac{\partial^2}{\partial \x^2} \phi_b(\x) &= \lambda_b \phi_b(\x), \hspace{1cm}  \x \in \Omega \\
    \phi_b(\x) &= 0, \hspace{2.1cm} \x \in \partial \Omega
\end{cases}
\end{equation}
for a compact set $\Omega \subset \mathbb{R}^D$. We see that this approximation has the same form as \ename~\ref{eq: approximation_idea} with $M=B$ and $\psi_m(\x) = \sqrt{s_m} \phi_m(\x)$. The spectral density has a closed form for many kernels, and the domain $\Omega$ can be selected so that the eigenvalues and eigenfunctions have one too. Functions $\psi_m(\x)$ are therefore easy to evaluate and the computation strategy described in \secname~\ref{s: basic_idea} can then be used. As an example, when $D=1$ and $\Omega = [-L, L]$ with $L > 0$, we have
\begin{equation}
    \begin{cases}
      \phi_b(\x) &= \frac{1}{\sqrt{L}} \sin \left( \frac{\pi b (\x + L)}{2L} \right)\\
   \lambda_b &= \left(\frac{\pi b}{2 L} \right)^2
\end{cases}.
\end{equation}
and it was proven in \citep{solin2019} that in this case $\lim_{L \rightarrow \infty} \left[\lim_{B \rightarrow \infty}  \tilde{k}(\x, \x') \right] = k(\x, \x')$ 
uniformly for any stationary $k$ that has a regular enough spectral density. For example for the EQ kernel (\ename~\ref{eq: eq_kernel}) $k_{\text{EQ}}(r) = \alpha^2 \exp \left(-\frac{r^2}{2\ell^2} \right)$, the spectral density is $S_{\text{EQ}}(\omega) = \alpha^2 \ell \sqrt{2 \pi} \exp \left(-\frac{\ell^2 \omega^2 }{2}  \right)$.

\subsection{Kernels for categorical variables}

\subsubsection{Eigendecomposition}

Let us study a kernel $k: \mathcal{X} \times \mathcal{X} \rightarrow \mathbb{V} \subset \mathbb{R}$, where $\mathcal{X}$ is a finite set of $C$ possible values (categories). We can encode these categories numerically as integers $\mathcal{X} = \{1, \ldots, C\}$. Because there are only $C^2$ possible input combinations $(v,w)$ for $k$, and therefore $\lvert \mathbb{V} \rvert \leq C^2$, we can list them in the $C \times C$ matrix $\bm{C}$ which has elements $\left[ \bm{C} \right]_{v,w} = k(v, w)$. The symmetric square matrix $\bm{C}$ has the orthogonal eigendecomposition
\begin{equation}
\label{eq: kcat_matrix_decomposition}
    \bm{C} = \bm{\Theta} \bm{D} \bm{\Theta}^{\top},
\end{equation}
where $\bm{D}$ is the diagonal matrix containing the eigenvalues $d_c$, $c=1, \ldots, C$ on the diagonal and $\bm{\Theta}$ has the $C$ eigenvectors as its columns. For each column $c$, we can define function $\varphi_c: \mathcal{X} \rightarrow \mathbb{R}$ so that $\varphi_c(v) = \left[\bm{\Theta} \right]_{v,c}$.  We see that
\begin{align}
\label{eq: kcat_decomposition}
    &k(v, w) = \left[ \bm{C} \right]_{v,w} = \left[ \bm{\Theta} \bm{D} \bm{\Theta}^{\top} \right]_{v,w} \\
    &= \sum_{c=1}^C d_c \left[\bm{\Theta} \right]_{v,c} \left[\bm{\Theta}\right]_{w,c} = \sum_{c=1}^C d_c \varphi_c(v) \varphi_c(w) \label{eq: ckd},
\end{align}
meaning that we have written $k$ in the form of \ename~\ref{eq: approximation_idea} with $M=C$ and $\psi_m(\x) = \sqrt{d_m} \varphi_m(\x)$. Note that this is an exact function decomposition for $k$ and not an approximation. The complexity of computing the eigendecomposition is $\mathcal{O}(C^3)$, but in typical applications
$C \ll N$ and this is not a bottleneck. Actually, for example
for the ZS kernel and other CS kernels, the eigenvalues have
a closed form and the corresponding eigenbasis is known (see next subsection). Furthermore, if $k$ does not depend on any hyperparameters, the eigendecomposition only needs to be done once before parameter inference. If it is of type $k(\x, \x') = \alpha^2 k_0(\x, \x')$ where $\alpha$ is the only parameter, the decomposition can obviously
be done just for $k_0$ which again has no parameters. Evaluating functions $\varphi_m$ is easy as it corresponds to just looking up a value from the matrix $\bm{\Theta}$.

\subsubsection{Compound symmetry kernels}
\label{s: compound_symmetry}
For a categorical variable $z$, the compound symmetry (CS) \citep{pinheiro2000} kernel function is
\begin{equation}
    k_{\text{CS}}(z, z' \mid \alpha, \rho) =
    \begin{cases}
       \alpha^2 \text{ if } z = z' \\
       \rho \ \ \text{ if } z \neq z'
    \end{cases},
\end{equation}
meaning that all within-group variances $\alpha^2 \geq 0$ are equal and all between-group covariances $\rho$ are too. For $z$ with $C$ different categories, we need the restriction 
$- \frac{\alpha^2}{C-1} \leq \rho \leq \alpha^2$
to ensure that $k_{\text{CS}}(z, z' \mid \alpha, \rho)$ is
positive semidefinite. By fixing $\rho = -\frac{\alpha^2}{C-1}$,
we recover $\alpha^2 k_{\text{ZS}}(z, z')$ as a special case.

The $C \times C$ matrix $\bm{C}$ that has
elements $\left\{\bm{C} \right\}_{v,w} = k_{\text{CS}}(v, w \mid \alpha, \rho)$ has eigenvalues
\begin{equation}
\begin{cases}
    d_1 &= \alpha^2 + (C-1)\rho \\
    d_c &=  \alpha^2 - \rho \hspace{2cm} c = 2, \dots, C
\end{cases}
\end{equation}
and the eigenvector corresponding to $d_1$ is the vector of ones $\bm{1}$, and the eigenspace of $d_c$, $c \in \{2, \ldots, C \}$, is the orthogonal complement $\bm{1}^{\perp}$ \citep{roustant2018}. The decomposition $\bm{C} = \bm{\Theta} \bm{D} \bm{\Theta}^{\top}$ can therefore be created for example so that the first column of $\bm{\Theta}$ is $\bm{1}$ normalized to unit length and remaining columns are normalized columns of the $C \times (C-1)$ Helmert contrast matrix \citep{chambers1992}. The diagonal matrix $\bm{D}$ has the eigenvalues $d_c$, $c=1, \ldots, C$ on its diagonal. For the ZS kernel, the eigenvalue $d_1$ is zero, meaning that we actually have a rank $C-1$ decomposition.

\subsection{Mixed-domain product kernels}
We now consider approximating a product kernel $k(\x, \x') = \prod_{i=1}^P k_i(\x, \x’)$ with $\tilde{k}(\x, \x') = \prod_{i=1}^P \tilde{k}_i(\x, \x’)$,
where for each $\tilde{k}_i$ we have an available decomposition
\begin{equation}
    \tilde{k}_i(\x, \x') = \sum_{m=1}^{M_i} \psi_{i,m}(\x) \psi_{i,m}(\x'),
\end{equation}
which might be an approximation or an exact decomposition of $k_i$. The total approximation is
\begin{align}
\label{eq: sumprod_decomposition}
    &\tilde{k}(\x, \x') = \prod_{i=1}^P \sum_{m=1}^{M_i} \psi_{i,m}(\x) \psi_{i,m}(\x') \\
    &= \sum_{m_1=1}^{M_1} \ldots \sum_{m_P=1}^{M_P} \psi_{m_1, \ldots, m_P}^*(\x) \psi_{m_1, \ldots, m_P}^*(\x') \label{eq: mdpk}
\end{align}
where $\psi^*_{m_1, \ldots, m_P}(\x) = \prod_{i=1}^P \psi_{i,m_i}(\x)$. We have now a representation of the product kernel in the form of \ename~\ref{eq: approximation_idea} with $M = \prod_{i=1}^P M_i $ sum terms. Note that since the individual kernels in the product kernel can be both categorical and continuous, \ename~\ref{eq: mdpk} provides a kernel representation for mixed-domain GPs with product kernels. Also note that $M$ grows exponentially with $P$.

An example with $P=2$ is the $k_{\text{EQ} \times \text{ZS}}((x, z), (x', z'))$  interaction kernel (see \ename~\ref{eq: eq_kernel}-\ref{eq: zs_kernel}), for which $M_1 = B$, $M_2 = C$, and
\begin{equation}
    \psi^*_{c, b}((x,z)) = \sqrt{s_b d_c}  \phi_b(x) \varphi_c(z).
\end{equation}

\subsection{Mixed kernels for longitudinal data}
In our framework, we consider mixed kernels $k: \mathcal{X} \times \mathcal{X} \rightarrow \mathbb{R}$, where $\mathcal{X}$ is a mixed space of both continuous and categorical dimensions, consisting of multiplication and addition so that
\begin{equation}\label{eq:mixeddomainlongdata}
    k(\x, \x') = \sum_{j=1}^J \left( \prod_{q=1}^{Q_j} k_{j,q}(\x, \x') \prod_{r=1}^{R_j} \kappa_{j,r}(\x, \x') \right),
\end{equation}
where each $k_{j,q}: \mathbb{R} \times \mathbb{R} \rightarrow \mathbb{R}$ is isotropic and depends only on one continuous dimension of $\x$ and each $\kappa_{j,r}: \{1, \ldots, C_{j,r}\} \rightarrow \mathbb{R}$ depends only on one categorical dimension of $\x$, which has $C_{j,r}$ different categories. For each $k_{j,q}$, we use the basis function approximation (\ename~\ref{eq: k_approx_trunc}) with $B_{j,q}$ basis functions and domain $\Omega = [-L_{j,q}, L_{j,q}]$, and for each $\kappa_{j,r}$ the exact decomposition (\ename~\ref{eq: ckd}). Using \ename~\ref{eq: sumprod_decomposition}, we can obtain an approximation for the mixed domain kernel for longitudinal data shown in \ename~\ref{eq:mixeddomainlongdata} that has the desired form $\tilde{k}(\x, \x') = \sum_{m=1}^M \psi_m(\x) \psi_m(\x')$ (\ename~\ref{eq: approximation_idea}) using  
\begin{equation}
   M = \sum_{j=1}^J \left( \prod_{q=1}^{Q_j} B_{j,q} \prod_{r=1}^{R_j} C_{j,r} \right) 
\end{equation}
terms. In each term, the function $\psi(\x)$ is a product of $Q_j$ factors $\sqrt{s} \phi(\x)$ and $R_j$ factors $\sqrt{d} \varphi(\x)$.

As an example, consider a commonly used additive longitudinal modeling approach where each additive term models the interaction effect for one continuous covariate and one categorical covariate, i.e., $Q_j = R_j = 1$ for each $j$. Assuming we use $B_{j,1} = B$ basis functions for all components, then the scalability is $\mathcal{O}(NM + M)$ where $M = B \sum_{j=1}^J C_{j,1}$. Further, if each categorical variable has $C$ many different values, then the scalability is $\mathcal{O}(NM + M)$, where $M = B J C$.

\subsection{Motivation and advice for base kernels}
The used base kernels $k_{j,q}$, $\kappa_{j,r}$ should be selected so that they reflect the assumed covariance structure over the dimension of $\textbf{x}$ in question. In this paper we use the exponentiated quadractic kernel since we have observed it to work well enough, and its spectral density has a known closed form expression. The practitioner who wishes to use a different kernel needs to derive or look up
its spectral density (\ename~\ref{eq: spectral_dens}). See for example \citep{rasmussen2006} for the formula for the spectral density of the Matern class of covariance functions. 

For categorical variables we have used the zero-sum kernel because it solves a non-identifiability problem of additive population-level and group-level effects, as the group level effects sum to zero when the ZS kernel is used. Furthermore, the eigendecomposition (\ename~\ref{eq: kcat_matrix_decomposition}) of any kernel of the compound symmetry class has a convenient closed form expression (\secname~\ref{s: compound_symmetry}) and the ZS kernel is a valid compound symmetry kernel by definition. For a different categorical kernel, one may need to
restrict its possible parameters so that positive semidefiniteness is ensured,
and compute the corresponding eigendecomposition. 

\section{Model reduction}

In many applications, a variety of potential explanatory variables are available and it is not known which should be included in the
model. If the only goal is prediction, it can be useful to include
all variables in the model. However, to reduce future measurement costs and improve interpretability, we often want to reduce the model as much as we can without losing predictive performance significantly. This is called \textit{minimal subset variable selection }\citep{pavone2020}. Compactly representing the data generating process is the goal also in \textit{descriptive modeling} \citep{shmueli2010}.

Here we compare two methods that are specifically suited for reducing the approximate additive GP models
of longitudinal data, but have not been extensively highlighted in the longitudinal modeling literature. The model reduction problem is to select a subset of the components $j \in \{1, \ldots, J\}$ in the sum $f = \sum_{j=1}^J f^{(j)}$.

\subsection{Additive variance decomposition}
\label{s: avd}
This method involves first fitting a full model $\mathcal{M}_{\text{full}}$ with all candidate components included. The amount of noise $p_{\text{noise}} \in (0,1)$ is estimated from it using a variance decomposition similar to
the Bayesian $R^2$ \citep{gelman2019}, and the remaining variance is decomposed between all the model components to obtain component relevances $\text{rel}_j \in (0,1-p_{\text{noise}})$, i.e., $p_{\text{noise}} + \sum_{j=1}^J \text{rel}_j = 1$. The relevances can be used to rank the components from most important to least important, and the total explained variance (cumulative relevance) given a subset $\mathcal{J} \subseteq \{1, \ldots, J\}$ of components is 
\begin{equation}
\label{eq: subset_rel}
    \text{rel}_{\mathcal{J}} = p_{\text{noise}} + \sum_{j\in \mathcal{J}} \text{rel}_j
\end{equation}
The final 
reduced model can be selected to be the smallest subset that has, for
example, $\text{rel}_{\mathcal{J}} \geq 0.95$.

\cite{timonen2021} demonstrated that the method works well for exact longitudinal GP models, outperforming a cross-validation based method \citep{cheng2019} in assessing the covariate relevances. However, the experiments were done on data sets with a relatively low number of potential explanatory variables, and the experiments with simulated data did not involve correlated predictors. 

\subsection{Projection predictive model selection}
\label{s: pp}
\subsubsection{Overview}
The projection predictive method \citep{goutis1998} is a model reduction method based on first fitting the best possible reference model $\mathcal{M}_{\text{ref}}$, which in our case will be the same as $\mathcal{M}_{\text{full}}$ but does not generally need to be. For a given submodel $\mathcal{M}_{\perp}$ containing a subset of the components, the submodel is fit by projecting the predictive distribution of $\mathcal{M}_{\text{ref}}$ onto $\mathcal{M}_{\perp}$.  This is done by minimizing KL divergence between the predictive distributions. In practice, the fit consists of MCMC draws and they have to be projected one by one. For models in the exponential family, this is equivalent to maximizing the likelihood of $\mathcal{M}_{\perp}$ when the data is replaced by the prediction given by $\mathcal{M}_{\text{ref}}$ \citep{pavone2020}. A good reference model therefore is able to filter noise in the data, and therefore stabilize the model reduction process. A good theoretical explanation of why the reference model is helpful is given by \cite{piironen2020}.

The projection predictive technique cannot directly give a relevance value for each model component, and has to be coupled with a search through the subspace of alternative submodels. This gives a path through the submodel space, and some rule is then needed for final model size determination so that the predictive distribution of $\mathcal{M}_{\text{sub}}$ is close enough to that of $\mathcal{M}_{\text{ref}}$ \citep{piironen2020}. An intuitive rule for deciding the reduced model size $k$ is to pick the smallest $k$ so that an estimated predictive utility, such as the expected log predictive density (ELPD), of the submodel is close to that of the reference model. Mathematically, this can be formulated as 
$\left \lvert \Delta_{\text{ELPD}}^k \right \rvert \leq 1$, where the relative difference in ELPD using model size $k$ is
\begin{equation}
\label{eq: elpd_rel_diff}
    \Delta_{\text{ELPD}}^k = \frac{\hat{u}^* - \hat{u}_k}{s^*_k},
\end{equation}
where $\hat{u}_k$ is the ELPD estimate for the submodel, $\hat{u}^*$ is the ELPD estimate for the reference model, and $s^*_k$ is the standard error for the reference model ELPD estimate. For other rules, see \cite{mclatchie2024}.

For various types of models \citep{catalina22a, mclatchie2024, piironen2017}, the projection predictive technique has been shown to be able to find relatively small subsets of predictors that still have a good predictive power. The technique is fairly mature for generalized linear models \citep{piironen2017}, but for complex types of models such as generalized linear mixed models and generalized additive (mixed) models (GAMs), it requires nontrivial tuning \citep{catalina22a}.

\subsubsection{Application for approximate GP models of longitudinal data}
The approximate additive GP models used in our experiments can be seen as GAMs, since the modeled function can be written as
\begin{equation}
\label{eq: gam_formulation}
    f(\bm{x}) = \sum_{j = 1}^J f^{(j)}(x_j, z_j) =  \sum_{j = 1}^J \sum_{b=1}^{B_j} {w_{j,b}} \phi_{j,b}(x_j)
\end{equation}
where $x_j$ is continuous\footnote{Note that $x_j$ in the formula can be the same variable for many $j$, if both its shared and category-specifc effects are modeled.}, $z_j$ is a categorical factor, and the parameter ${w_{j,b}}$ can be specific to group $z_j$. We note that shared effect terms are actually of the form $f^{(j)}(x_j)$ and some terms may be only a category-specific offset, having the form $f^{(j)}(z_j)$.

GAM parameters are typically fitted by penalizing the likelihood with a wiggliness penalty term to avoid overfitting and \cite{catalina22a} used similar penalization when 
applying the projection predictive method for GAMs. For exponential family models, projecting a reference model prediction $f^*$ to submodel $\mathcal{M}_{\perp}$ corresponds to finding coefficients $w_{j,b}$ of the submodel that maximize the likelihood $\mathcal{L}(f^* \mid \bm{\theta})$. However, as this optimization problem is not identifiable, we instead maximize a penalized log likelihood
\begin{equation}\label{eq:penal-log-likelihood}
    \mathcal{L}_{\gamma}(f^* \mid \bm{\theta}) = \log \mathcal{L}(f^* \mid \bm{\theta}) + \sum_{j \in \mathcal{M}_{\perp}} \gamma_j \int_{\Omega} \left[\frac{\partial^2}{\partial x_j^2}f^{(j)}(x_j) \right]^2 \text{d}x_j ,
\end{equation}
where we have assumed $f^{(j)}(x_j, z_j) = f^{(j)}(x_j)$ for simplicity. For how the categorical effects affect the penalization, we refer to \cite{wood2017}, since we use the \texttt{gam()} function of the \texttt{mgcv} R package to solve the penalized optimization problem that performs the projection. This is achieved by defining a custom \texttt{smooth} that corresponds to the Hilbert space GP basis functions, and implementing the corresponding penalty for it. Submodel terms of the form 
\begin{itemize}
    \item $f^{(j)}(x_j)$ correspond to \texttt{s(x\_j)}
    \item $f^{(j)}(z_j)$ correspond to \texttt{z\_j}
    \item $f^{(j)}(x_j, z_j)$ correspond to \texttt{s(x\_j, by = z\_j) + z\_j}
\end{itemize}
in the \texttt{R} formulas that we use with \texttt{gam()}.  We use the \texttt{"GCV.Cp"} option to select the smoothing parameters~$\gamma_j$.

Note that since the reference model fit consists of MCMC draws and the projection is done draw-by-draw, the penalised log likelihood optimization problem needs to be solved for the reference model predictions $f^*$ computed using each draw. We use a subset of draws from the posterior to speed up the projection and be able to scale to larger sets of alternative submodels.

\subsubsection{Penalization for Hilbert space basis functions}
We derive here the second derivative-based wiggliness penalty for the Hilbert space basis functions of the
EQ kernel. Assume that the additive submodel has a term
$f_j$ modeled as an approximate GP (\ename~\ref{eq: k_approx_trunc}) with basis functions $\phi_{j,b}(x) = \frac{1}{\sqrt{L}} \sin \left( \frac{\pi b (x + L)}{2L} \right)$, $b = 1, \ldots, B$ and domain $\Omega = [-L, L]$. The integral in the penalization term $j$ in \ename~\ref{eq:penal-log-likelihood} becomes
\begin{align}
\int_{\Omega} \left[f''(x) \right]^2 dx &= \int_{\Omega} \left[ \sum_{b=1}^B \sqrt{s_{b}} \phi_{b}''(x) \right]^2 \text{d}x,
\end{align}
where we have dropped the subscript $j$ and superscript $(j)$ for readability. We have
\begin{equation}
    \phi_{b}''(x) = - \frac{(\pi b)^2}{(2L)^2} \cdot \phi_{b}(x) 
\end{equation}
and can write
\begin{align}
\int_{\Omega} \left[f''(x) \right]^2 \text{d}x = \bm{w}^\top \bm{S} \bm{w},
\end{align}
where
\begin{align}
\left[ \bm{S} \right]_{l,m} &= \int_{\Omega} \phi_{l}''(x) \phi_{m}''(x) \text{d}x = 0
\end{align}
for $l \neq m$ and
\begin{equation}
    \left[ \bm{S} \right]_{b,b} = \left(\frac{\pi b}{2 L} \right)^4
\end{equation}
on the diagonal. See derivation in \secname~S3.4. This means that basis functions with a higher frequency in the sine wave (larger $b$) are penalized more.

\section{Results}
We confirm the scalability and accuracy of the presented approach
using experiments with simulated and real data. Code for reproducing the experiments is  publicly available \footnote{\url{https://github.com/jtimonen/scalable-mixed-domain-GPs}}. In the same way as the earlier \texttt{lgpr} package \citep{timonen2021}, the new R package \texttt{lgpr2} has a user-friendly interface based on the formula syntax in R and fits the described longitudinal approximate GP models using Stan \citep{carpenter2017}. 

In all experiments, used the same number of basis functions $B$ for
each approximate continuous kernel, and for all approximate
components the same domain scaling factor $c$, which is defined so that $\Omega=[-L,L]$ with $L$ being $c$ times the half-range of the continuous covariate of the approximated kernel \citep{riutortmayol2020}. Additional experiment details are documented in \secname~S1.1.

\label{s: results}

\begin{figure}
\includegraphics[width=0.85\linewidth]{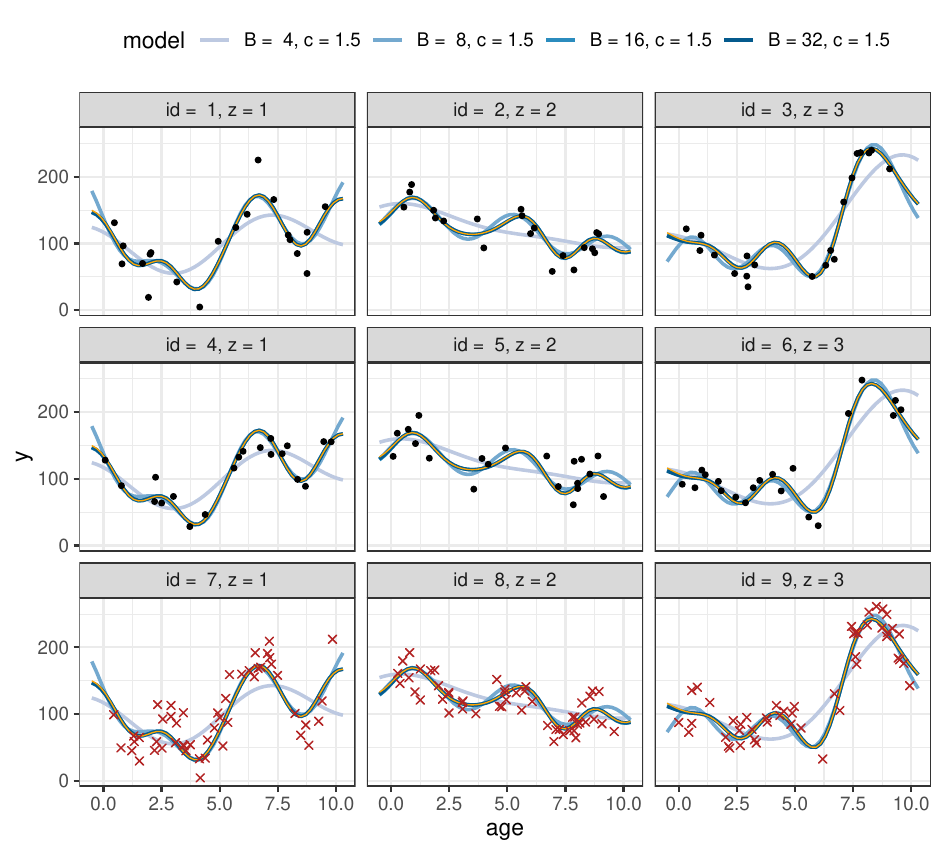}
\caption{The posterior predictive mean for four different
approximate models in one replication of Experiment 1. The models differ only based on the number of basis functions $B$ and domain scaling factor $c$. The yellow line 
shows the posterior predictive mean for the corresponding
exact GP model as a reference. Black dots are training
data and red crosses test data.}
\label{fig: simstudy_predictions}
\end{figure}

\subsection{Experiment 1: Simulation study}

In the first experiment we create simulated longitudinal data consisting of categorical variables $id$ and $z$, and a continuous variable $age$.
We create data with 9 individuals, where individuals with $id \in \{1,4,7\}$ belong
to group $z=1$, individuals with $id \in \{2,5,8\}$ belong
to group $z=2$ and individuals $id \in \{3,6,9\}$ belong
to group $z=3$. For each individual 1-6, we create $\frac{N_{\text{train}}}{6}$ observations at time points where $age$ is drawn uniformly from the interval $[0, 10]$, and $N_{\text{train}}$ is varied as $\{60,90,120,150,180,210\}$. For individuals 7-9, $\frac{N_{\text{test}}}{3}$ observations are created similarly, with
$N_{\text{test}} = 150$.

\begin{figure}
\centering
\includegraphics[width=0.76\linewidth]{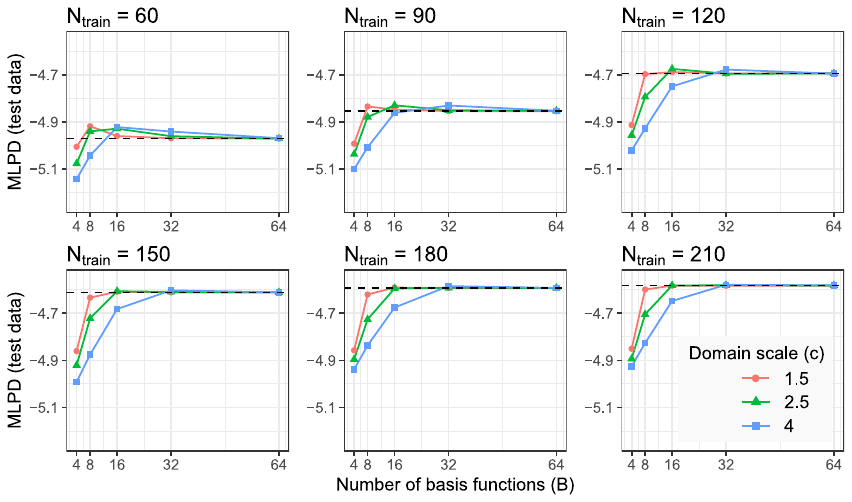}
\caption{Mean log predictive density for test data
in Experiment 1 for varying training data set sizes $N_{\text{train}}$ and domain scales $c$. Black dashed line corresponds to
the exact model. Results are averages over 30
replications of the experiment.} 
\label{fig: simstudy_twocomp_mlpd}
\end{figure}

We consider an additive GP model 
$f = f^{(1)}(age) + f^{(2)}(age, z)$ with kernels
\begin{equation}
\label{eq:simulationmodel1}
    \begin{cases}
       k_1 &=  \alpha_1^2 k_{\text{EQ}}(age, age'  \mid \ell_1) \\
       k_2 &= \alpha_2^2 k_{\text{ZS}}(z, z) k_{\text{EQ}}(age, age'  \mid \ell_2)
    \end{cases}
\end{equation}
and we simulate a realization of $\bm{f} \in  \mathbb{R}^{N_{\text{test}} + N_{\text{train}}}$ data using $\alpha_1=\alpha_2=1$, $\ell_1=2$ and
$\ell_2=1$. We then generate a response variable measurements 
$\bm{y} = 100 + 10  \cdot (\bm{f} + \bm{\epsilon})$, where $\epsilon_i \sim \mathcal{N}(0, 0.5^2)$ and the realization $\bm{f}$ represents the ground truth signal.

Data from individuals 1-6 is used in training, while data from individuals 7-9 is left for testing. Using the training data, we fit
an exact and approximate model with the correct covariance structure from Eq.~\ref{eq:simulationmodel1} using the Gaussian 
likelihood model. Exact model is fitted with \texttt{lgpr} \citep{timonen2021},
which also uses Stan for MCMC. The exact model utilizes
the marginalization approach for GPs (\secname~S3.2-S3.3), since Gaussian observation model is specified.

\begin{figure}
\centering
\includegraphics[width=0.8\linewidth]{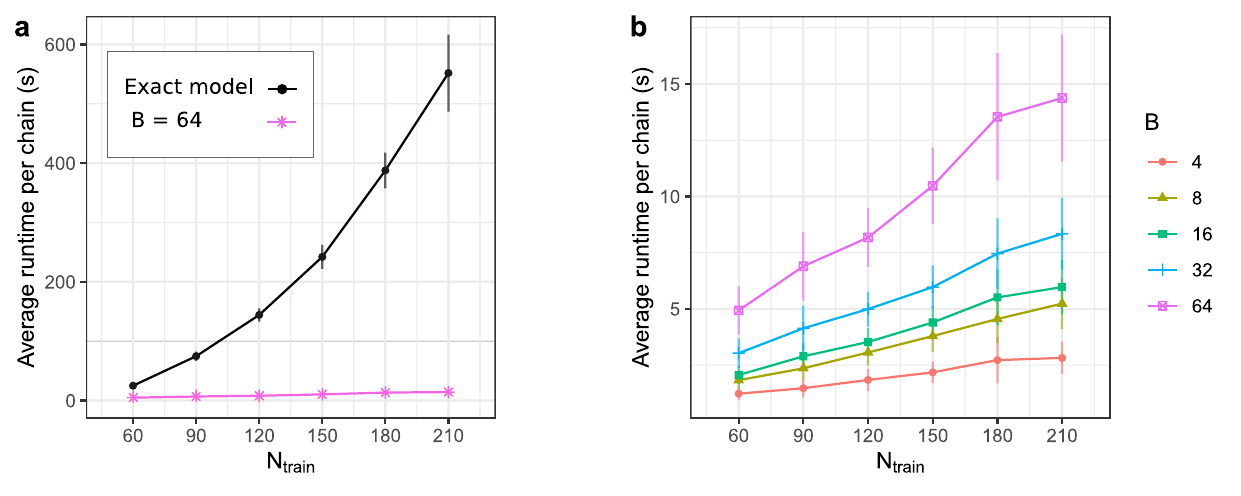}
\caption{Runtimes of fitting the exact and approximate
models in Experiment 1. \textbf{a)} Exact model vs. approximation with $B=64$.
\textbf{b)} Approximations using different values for $B$.
The markers show the average
time taken to run a chain, when a total of
$30 \times 4$ MCMC chains were run
for 2000 iterations each. The vertical error bars
show $\pm$ one standard deviation (not shown for approximate model in panel \textbf{a}. Note the smaller y-axis scale in panel \textbf{b}. We see empirically that the runtime of the approximate
model scales linearly in both $N$ and $B$.} 
\label{fig: simstudy_twocomp_runtimes}
\end{figure}

\fname~\ref{fig: simstudy_predictions} shows the posterior
predictive mean of the exact model and different approximate
models with $c=1.5$, using $N_{\text{train}} = 150$. We see that with $B=16$ and $B=32$ the mean predictions are indistinguishable from the exact model. We fit the approximate model using different values of $B$ and $c$, and repeat the experiment using different values for $N_{\text{train}}$.
Results in \fname~\ref{fig: simstudy_twocomp_runtimes} validate
empirically that the runtime scales linearly as a function of both $N_{\text{train}}$ and $B$. 

We compute the mean log predicive density (MLPD), at test points 
$\bm{y}^{*} \in \mathbb{R}^{150}$ (see \secname~S3.1.2 for details about
out-of-sample prediction and MLPD).  Results in \fname~\ref{fig: simstudy_twocomp_mlpd} show that the MLPD of the approximate
model approaches the exact model as $B$ grows. It is seen that with
small data sizes and small $B$, the predictive performance can actually be better than that of the exact model, possibly because the coarser
approximation is a simpler model that generalizes better in this case.

\subsection{Experiment 2: Canadian weather data}
We analyze data that consists of yearly average 
temperature measurements in 35 Canadian weather stations
\citep{ramsay2005}. There is a total of $N = 35 \times 365 = 12775$
data points, which are daily temperatures at the 35 locations, averaged over the years 1960-1994. We fit an additive GP model 
$f = f^{(1)}(day) + f^{(2)}(day, region) + f^{(3)}(day, station)$,
with Gaussian likelihood using the EQ kernel for $f^{(1)}$ and the product EQ$\times$ZS kernel for $f^{(2)}$ and $f^{(3)}$.

\begin{figure}
\includegraphics[width=0.8\linewidth]{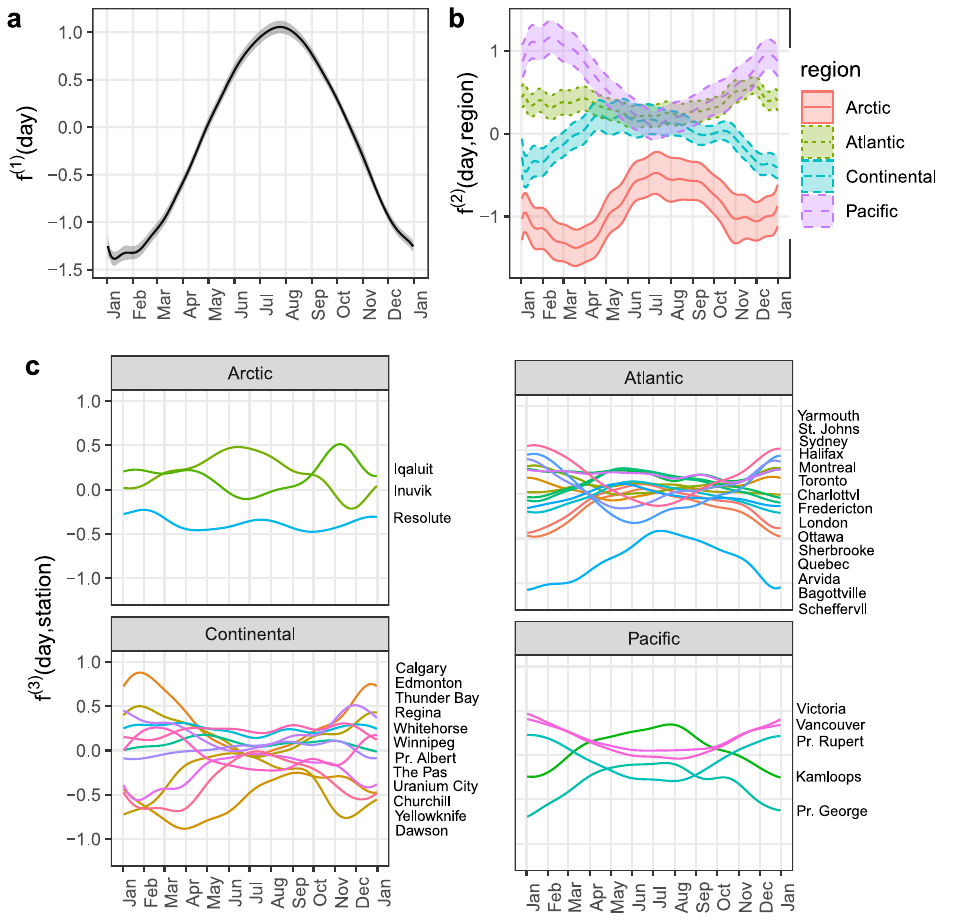}
\caption{Results for the Canadian weather data experiment with
$B=32$ and $c=1.5$. Panels \textbf{a}-\textbf{c} show the marginal posterior distribution
of each of the three components $f^{(1)}$, $f^{(2)}$ and $f^{(3)}$
(mean $\pm$ two times standard deviation). Standard deviation is
not shown for $f^{(3)}$ for clarity. The functions are on the
standardized scale (response variable normalized to zero mean and
unit variance). We see for example that the regions tend to have larger differences during winter. Notice that due to the specified categorical correlation structure, the effects of each region ($f^{(2)}$) sum to zero at each time point, and so do also the effects specific to each station ($f^{(3)}$). The sum-to-zero constraint disentangles the region or station specific time effects from the shared time effects.}
\label{fig: temperature_components}
\end{figure}

We use domain scaling factor $c=1.5$
for all components and run 4 MCMC chains in parallel using 4 CPU cores on a cluster. This was repeated with different values of $B=8,12,16,24,32$,
where $B$ is the number of basis functions for each component. Total
runtimes for fitting the models were $4.10, 7.24, 7.47, 14.18$ and
$18.76$ hours, respectively. The posterior distributions of each
model component with $B=32$ are in \fname~\ref{fig: temperature_components}. The posterior predictive distribution
for each station separately is visualized in \fname~S1. Note that this data set with $12755$ data points is large enough to make MCMC sampling for exact additive GP model fitting prohibitively slow. 

\begin{figure}
\includegraphics[width=0.95\linewidth]{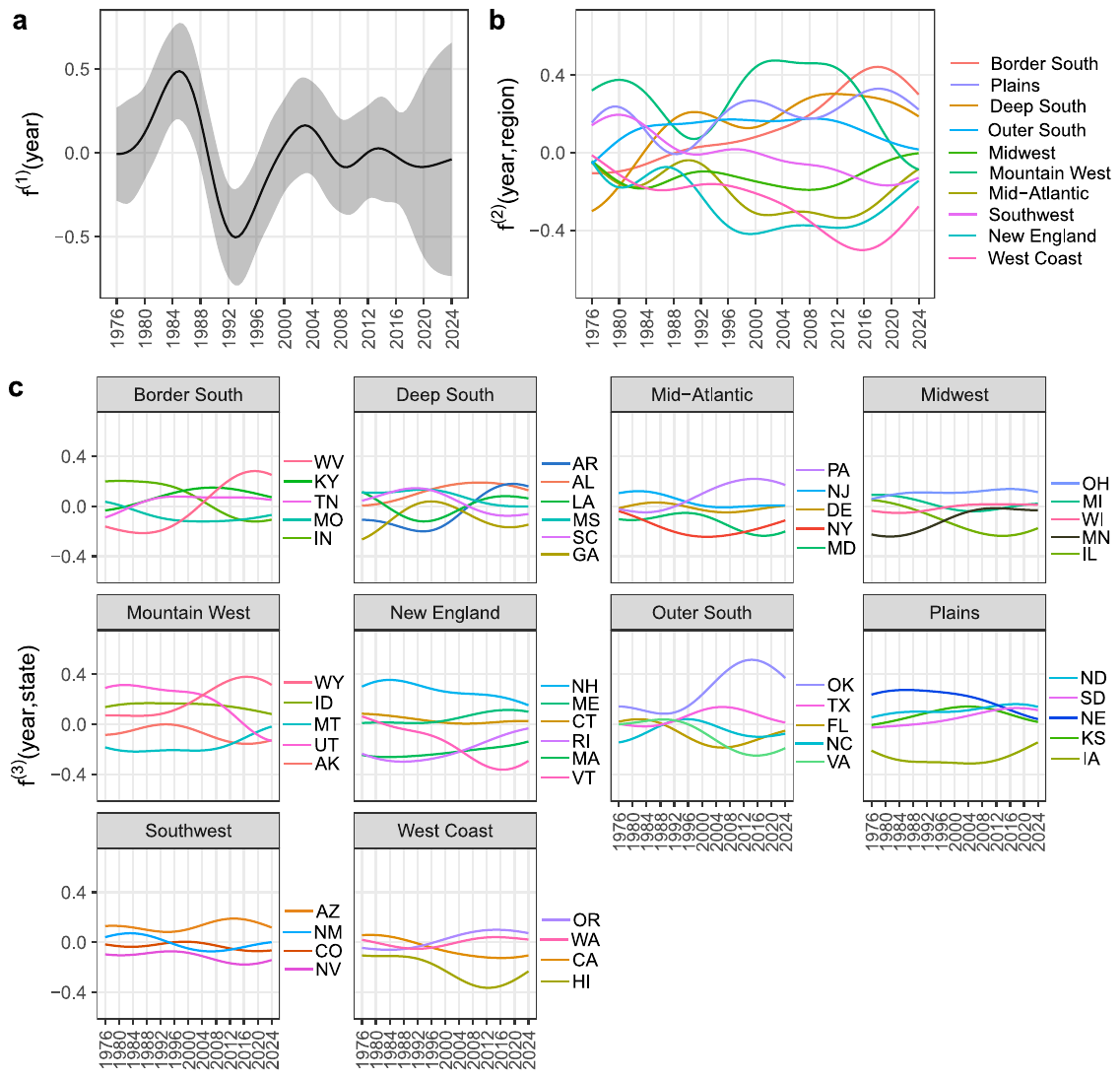}
\caption{Results for the US election prediction experiment with $B=24$ and $c=2.0$. Figures \textbf{a)}-\textbf{c)} show the marginal posterior distribution
of each of the three components $f^{(1)}$, $f^{(2)}$ and $f^{(3)}$
(mean $\pm$ two times standard deviation). Standard deviation is
not shown for $f^{(2)}$ and $f^{(3)}$ for clarity.}
\label{fig: election_components}
\end{figure}

\subsection{Experiment 3: US presidential election prediction}
In this example we demonstrate a beta-binomial observation model
and model the vote share of the Republican Party in each state in US presidential elections. By two-party vote share we mean proportion of votes cast to the Republican candidate divided by the sum of votes to both the Republican and Democratic candidates\footnote{Data is from \citep{mit2017}.}. Following \cite{trangucci2017}, Washington
DC is excluded from the analysis. We use data from the 1976-2016 elections as training data, meaning that $N = 50 \times 11 = 550$.

We fit an additive GP model 
$f = f^{(1)}(year) + f^{(2)}(year, region) + f^{(3)}(year, state)$,
with beta-binomial observation model using the EQ kernel for $f^{(1)}$ and the product EQ$\times$ZS kernel for $f^{(2)}$ and $f^{(3)}$. The observation model is
\begin{equation}
    \text{y}^{R} \sim \text{Beta-Binomial}(\text{y}^{R}+\text{y}^{D}, a, b),
\end{equation}
where $y^R, y^D$ are the number of votes for the Republican and Democratic parties, 
respectively, $a = \rho(\gamma^{-1} - 1)$, $b = (1-\rho)(\gamma^{-1} - 1)$, and
$\rho = \text{inv-logit}(f + w_0)$. We use a $\mathcal{N}(0, 0.5^2)$ prior for
the intercept $w_0$ and a Log-Normal(1,1) prior for the $\gamma$ parameter.

The posterior distributions of each
model component are in \fname~\ref{fig: election_components}. See also \fname~S2, where we have
also visualized the data from 2020 election to validate that the
model predicts well into the future. Fitting a model with $B=24$ and $c=2.0$ on a laptop computer\footnote{2018 MacBook Pro, 3 GHz Quad-Core Intel i5 CPU} running 4 MCMC chains in parallel took approximately 18 minutes. It is possible to fit an exact GP model to a dataset of this size, but we estimate it would take more than a day on the same machine.


\subsection{Experiment 4: Model reduction with simulated data}
We use the \texttt{lgpr} package to simulate data from
an additive GP model with a shared {\em age} effect (nonlinear), {\em id}-specific effect (baseline offset),
a category-specific effect of a 2-level categorical variable $z$ (nonlinear), a category-specific effect of a 3-level categorical variable $r$ (nonlinear), 
a shared effect of a continuous variable $x$ (nearly linear), and
a shared effect of a continuous variable $w$ (nearly linear). Furthermore, we add
$U=16$ continuous nuisance variables $x_1, ..., x_U$ and $U=16$ 2-level categorical
nuisance variables $z_1, \ldots, z_U$. While $x, w$ are generated independently, the continuous nuisance variables are drawn
from a multivariate Gaussian distribution so that their correlation with $x$ is $\rho = 0.85$. The categorical nuisance variables are created from $z$ so that their category is flipped for $\frac{1}{3}$ of the individuals, in order to induce correlation also between the categorical variables and thus make the model reduction task more difficult. The number of individuals is $50$, and with $16$ observations each, the data set size is $N = 800$.

The signal in the true model to generate data is  
\begin{equation}
    f_{\text{true}} = f^{(1)}(id) + f^{(2)}(age) + f^{(3)}(age, z) + f^{(4)}(age, r) + f^{(5)}(x) + f^{(6)}(w)
\end{equation}
and Gaussian noise is added so that signal-to-noise ratio is $\text{SNR} = \{0.1, 0.25, 0.5\}$. The fitted reference model $\mathcal{M}_{\text{ref}}$ is
\begin{equation}
    f_{\text{ref}} = f_{\text{true}} + \sum_{u = 1}^U f_x^{(u)}(x) + \sum_{u = 1}^U f_z^{(u)}(age, z_u)
\end{equation}
and we use the EQ kernel to model the shared effects and the EQ$\times$ZS product kernel for category-specific {\em age} effects, while the effect of $id$ is a subject-specific offset. The total number of terms is therefore $6 + 2 U = 38$. We compare methods to reduce this model to the simplest possible, i.e.\ ideally the one with the ''true'' relevant covariates $id, age, x, z, w, r$, or a small model with
equal predictive performance. When performing a forward search, we always assume that $f^{(1)}(id)$ and $f^{(2)}(age)$ are in the model due to their special nature in longitudinal modeling. 
 Additional details on the experiment setup are in \secname~S1.2.
 
We compare two search heuristics for the projection predictive method; a greedy forward search based on KL divergence against the reference model, and a pre-specificed search path given by the component
relevances determined via the additive variance decomposition (AVD, \secname~\ref{s: avd}). To speed up the search, we project only a subsample of 30 posterior draws during forward search, and rank the submodels based on the mean pointwise KL divergence
between the projected predictive distribution of the submodel and reference model. After every forward search step $k$, we project 100 draws to compute the expected projected log predictive density (ELPD) estimate $\hat{u}_k$ for the chosen submodel, approximated
using Pareto-smoothed importance sampling leave-one-out cross validation (PSIS-LOO) \citep{vehtari2017}. ELPD for the reference model, $\hat{u}^*$, is computed using PSIS-LOO with all posterior draws. Both stepwise methods are run until the model has $k=8$ terms. 

\begin{figure}
\includegraphics[width=\linewidth]{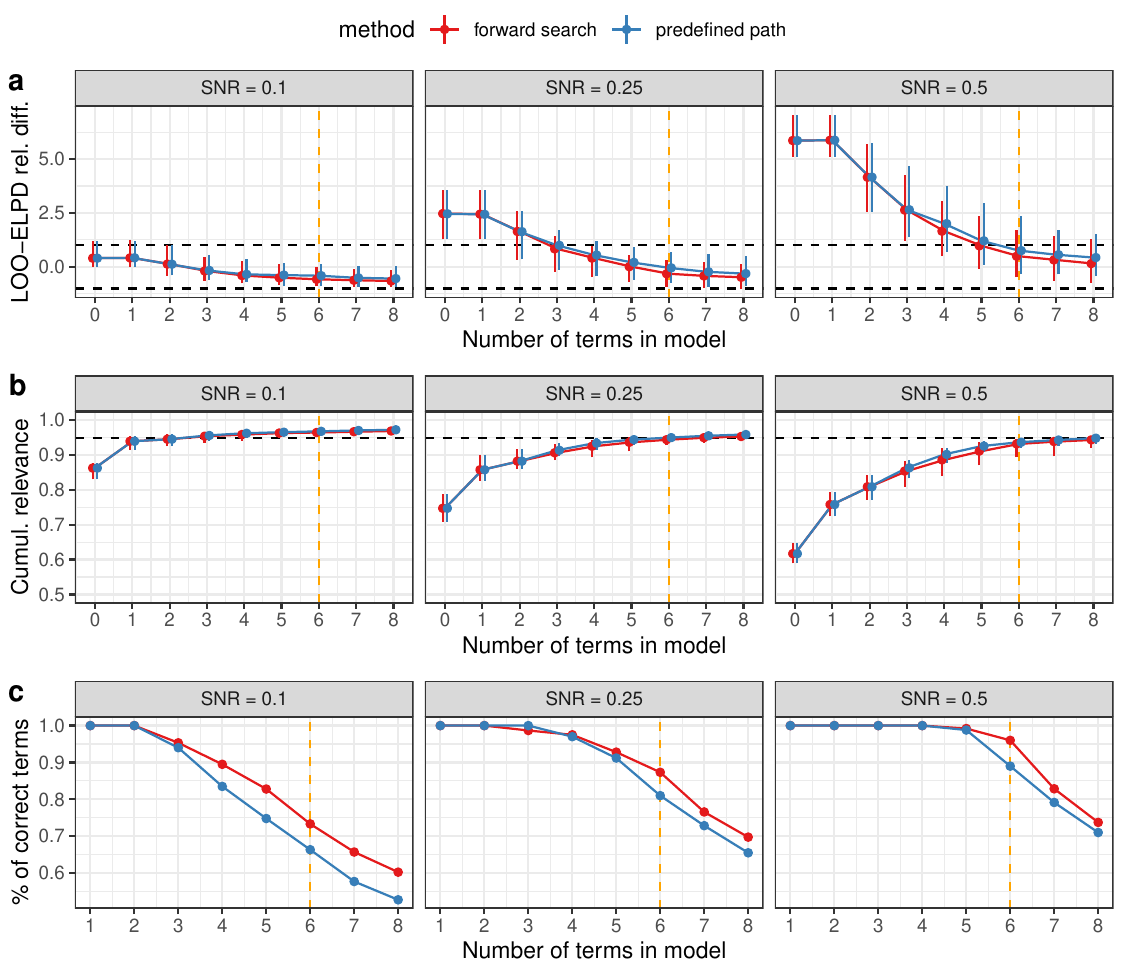}
\caption{Model reduction comparison experiment, with different values for the signal-to-noise ratio $\text{SNR} = \{0.1, 0.25, 0.5\}$. The orange dashed vertical line shows the true model size used to generate the data.  \textbf{a)} The estimated relative difference $\Delta_{\text{ELPD}}^k$ in predictive performance (lower is better) shows that the greedy forward search is able to find on average slightly better models with a given number of variables $k$ than the predefined path based on the componentwise relevances obtained via the additive variance decomposition (AVD). The black horizontal dashed lines represent the model size selection rule $\left \lvert \Delta_{\text{ELPD}}^k \right \rvert\leq 1$, and we see that the model size needed to satisfy it depends on the SNR. \textbf{b)} Cumulative relevance (explained variation) of the submodel with a given number of terms, given by AVD. The black horizontal dashed lines represent the model size selection threshold $0.95$, and we see that the model size needed to satisfy it also depends on the SNR. The dots in panels \textbf{a}-\textbf{b} show the median and the error bars show 90\% central credible intervals over the 50 repetitions. \textbf{c)} Proportion of ''true'' terms in the model at each step of the model search path. Shown are averages over the
50 repeated simulation experiments.}
\label{fig: comparison-pred-results}
\end{figure}

\fname~\ref{fig: comparison-pred-results}a shows, as a function of model size $k$, the estimated relative difference in ELPD (\ename~\ref{eq: elpd_rel_diff}). We see that the difference between the two search heuristics is very small, but the forward search can achieve a slightly better ELPD on average at given $k$. With $\text{SNR} = 0.1$, the data is so noisy that already the empty model has similar predictive performance as the reference model. With $\text{SNR} = 0.25$, three terms are usually required and with $\text{SNR} = 0.5$, five to six terms are required. 

In \fname~\ref{fig: comparison-pred-results}b,
we provide as comparison the cumulative relevance (\ename~\ref{eq: subset_rel}) given by AVD at each step $k$, using the same two search heuristics for the submodel. Plotted with the black horizontal dashed line is a possible threshold $0.95$, used in \cite{timonen2021}, which seems to result in larger model sizes than projection predictive method, difference being clearest in the $\text{SNR} = 0.1$ case where three to four variables are usually needed.

In \fname~\ref{fig: comparison-pred-results}c, we compare the proportion of ''true'' terms in the model by each method at each step of the search path. The forward search path seems to perform better choices on average. To shed light on the types of wrong selections, we plot the distribution of the selected term at steps $k=3,4,5,6$ (i.e. after the {\em id} and {\em age} terms that are fixed in the model at $k=1,2$) in \fname~S3. We see for example that selecting categorical nuisance variables into the model is more common than selecting continuous nuisance variables in this experiment.

In Table~S1, we report the average runtimes for different parts of performing the model reduction workflow. We see that running the full forward search for 8 steps takes a similar amount of time as fitting the reference model, and the search using the predefined search path based on AVD takes an order of magnitude less time, as computing the AVD has a negligible runtime. We note that in practice the search would be stopped as the stopping condition for model size is met, which usually occurs earlier. 

An additional experiment with a larger data set and different number of true variables is described in \secname~S2.1.

\section{Limitations}
 The approximation method that we use requires that the continuous kernels are stationary. Non-stationary effects can still be modeled by applying a warping on the input first, and then using a stationary kernel (see for example \cite{snoek2014}). Another limitation is that we use kernels that are composed of sums and products of base kernels. It is possible to construct a mixed-domain kernel that does not fit into this family, as is done in \secname~S2.2, where we study one such kernel. However, we find this example rather artificial, and in reality additive kernels with low-dimensional base kernels have demonstrated good predictive performance \citep{lu22}, and have the advantage of increased model interpretability.

 In \secname~\ref{s: results}, the largest data size that was analyzed was $N = 12775$ observations. In \secname~S2.3, we push the modeling framework further to study the practical scalability limit. We find estimate that with our implementation, it is viable to analyze data sets on the order of $N=10^5$ when the total number of basis functions $M$ is on the order of hundreds (see \secname~\ref{s: basic_idea}). This practical limit arises largely due to using full Bayesian inference with MCMC, as Stan does not scale well to millions of observations.

\section{Conclusion}

Gaussian processes offer an attractive framework for specifying flexible models using a 
kernel language. The computational cost of their exact inference however limits possible applications to small data sets, and the complexity of alternative model spaces poses additional limits for the scalability of a useful model building workflow. Our scalable framework opens up a rich
class of GP models to be used in large scale applications of various fields of science as the computational complexity is linear with respect to data size. We have presented a scalable approximation scheme for mixed-domain covariance functions,
and demonstrated its use in the context of Bayesian GP regression. However, it can
also be applied in GP applications where the kernel hyperparameters are optimized using
a marginal likelihood criterion.

Based on the model reduction experiments, the projection based full forward search seems to be able to rank the variables in a better order based on whether they were actually in the data-generating model. However, we note that the practical relevance of the ''false'' variables that are correlated with the ''true'' variables is not zero since they retain some of predictive information of the ''true'' variables. The additive variance decomposition (AVD) method seems to perform remarkably well in selecting small models that have a good predictive performance even in challenging settings and in the presence of a multitude of collinear predictors. It is a viable candidate for ranking the variables in applications where performing a full forward search is prohibitively slow due to the larger number of alternative models at each step, and also useful for determining the final model size. One reason for why the AVD works well for these longitudinal models is the fact that in the full model we use the zero-sum categorical kernel in interaction terms with age to disentangle the effect of the categorical variable on the explained variance from that of the shared effect of age.

Note that in the model reduction experiments we used approximate estimation of the ELPD and did not test the performance on actual hold out data. 
In practice it is important to interpret the projected posterior as explained in \cite{mclatchie2024}, to assess whether it can actually be used for out-of-sample prediction in addition to just model ranking during the model space search. A safe alternative is to refit the final found submodel using its own posterior instead of the projected posterior for final evaluation or prediction. In addition, it can be useful to diagnose the model search path using cross-validation over the entire model search if signs of over-optimism are detected \citep{mclatchie2024}. One sign is the ELPD of the submodel exceeding the ELPD of the reference model, which we did not see in our experiments, and therefore did not cross-validate the search path due to increased computational demands. The final model size determination is always inherently dependent on some threshold rule and setting the optimal threshold depends on factors such as the amount of noise in the data, as seen in our experiments (see also \secname~S2). For the models discussed in this paper, it seems beneficial to look at both the relative difference in ELPD and cumulative relevance, and whether these metrics plateau rather than whether they exceed a given threshold.


\begin{acks}[Acknowledgments]
We thank Aki Vehtari and Gleb Tikhonov for useful comments on early versions of this manuscript, and acknowledge the computational resources provided by Aalto Science-IT, Finland. We wish to also thank anonymous reviewers for their constructive comments.
\end{acks}

\begin{funding}
This work was supported by the Academy of Finland and Bayer Oy.
\end{funding}

\begin{supplement}
\stitle{Scalable mixed-domain Gaussian process modeling and model reduction for longitudinal data: Supplementary material}
\sdescription{Sections S1-S3, Figures S1-S7, Table S1.}
\end{supplement}

\bibliographystyle{ba}
\bibliography{main}

\begin{thebibliography}{45}
\newcommand{\enquote}[1]{``#1''}
\expandafter\ifx\csname natexlab\endcsname\relax\def\natexlab#1{#1}\fi
\expandafter\ifx\csname url\endcsname\relax
  \def\url#1{{\tt #1}}\fi
\expandafter\ifx\csname urlprefix\endcsname\relax\def\urlprefix{URL }\fi
\ifx\endbibitem\undefined \let\endbibitem\relax\fi

\bibitem[{Cao et~al.(2015)Cao, Brubaker, Fleet, and Hertzmann}]{cao2015}
Cao, Y., Brubaker, M.~A., Fleet, D.~J., and Hertzmann, A. (2015).
\newblock \enquote{Efficient Optimization for Sparse {G}aussian Process
  Regression.}
\newblock {\em IEEE Transactions on Pattern Analysis and Machine
  Intelligence\/}, 37(12): 2415--2427.
\endbibitem

\bibitem[{Carpenter et~al.(2017)Carpenter, Gelman, Hoffman, Lee, Goodrich,
  Betancourt, Brubaker, Guo, Li, and Riddell}]{carpenter2017}
Carpenter, B., Gelman, A., Hoffman, M.~D., Lee, D., Goodrich, B., Betancourt,
  M., Brubaker, M., Guo, J., Li, P., and Riddell, A. (2017).
\newblock \enquote{Stan: A Probabilistic Programming Language.}
\newblock {\em Journal of Statistical Software\/}, 76(1): 1--32.
\endbibitem

\bibitem[{Catalina et~al.(2022)Catalina, B\"urkner, and Vehtari}]{catalina22a}
Catalina, A., B\"urkner, P.-C., and Vehtari, A. (2022).
\newblock \enquote{Projection Predictive Inference for Generalized Linear and
  Additive Multilevel Models.}
\newblock In {\em Proceedings of The 25th International Conference on
  Artificial Intelligence and Statistics\/}.
\endbibitem

\bibitem[{Chambers and Hastie(1992)}]{chambers1992}
Chambers, J.~M. and Hastie, T.~J. (1992).
\newblock {\em Statistical Models in {S}\/}.
\newblock London: Chapman \& Hall, 1st edition.
\endbibitem

\bibitem[{Cheng et~al.(2019)Cheng, Ramchandran, Vatanen, Lietzen, Lahesmaa,
  Vehtari, and L{\"a}hdesm{\"a}ki}]{cheng2019}
Cheng, L., Ramchandran, S., Vatanen, T., Lietzen, N., Lahesmaa, R., Vehtari,
  A., and L{\"a}hdesm{\"a}ki, H. (2019).
\newblock \enquote{An additive {G}aussian process regression model for
  interpretable non-parametric analysis of longitudinal data.}
\newblock {\em Nature Communications\/}, 10.
\endbibitem

\bibitem[{Chung et~al.(2020)Chung, Kim, Lee, Kim, Hwang, and Yang}]{chung2020}
Chung, I., Kim, S., Lee, J., Kim, K.~J., Hwang, S.~J., and Yang, E. (2020).
\newblock \enquote{Deep Mixed Effect Model using {G}aussian Processes: A
  Personalized and Reliable Prediction for Healthcare.}
\newblock {\em The Thirty-Fourth AAAI Conference on Artificial Intelligence
  (AAAI-20)\/}.
\endbibitem

\bibitem[{Deng et~al.(2017)Deng, Lin, Liu, and Rowe}]{deng2017}
Deng, X., Lin, C.~D., Liu, K.-W., and Rowe, R.~K. (2017).
\newblock \enquote{Additive {G}aussian Process for Computer Models With
  Qualitative and Quantitative Factors.}
\newblock {\em Technometrics\/}, 59(3): 283--292.
\endbibitem

\bibitem[{Fortuin et~al.(2021)Fortuin, Dresdner, Strathmann, and
  Rätsch}]{fortuin2021}
Fortuin, V., Dresdner, G., Strathmann, H., and Rätsch, G. (2021).
\newblock \enquote{Sparse {G}aussian Processes on Discrete Domains.}
\newblock {\em IEEE Access\/}, 9: 76750--76758.
\endbibitem

\bibitem[{Garrido-Merchán and Hernández-Lobato(2020)}]{garridomerchan2018}
Garrido-Merchán, E.~C. and Hernández-Lobato, D. (2020).
\newblock \enquote{Dealing with categorical and integer-valued variables in
  {B}ayesian Optimization with {G}aussian processes.}
\newblock {\em Neurocomputing\/}, 380: 20--35.
\endbibitem

\bibitem[{Gelman et~al.(2019)Gelman, Goodrich, Gabry, and Vehtari}]{gelman2019}
Gelman, A., Goodrich, B., Gabry, J., and Vehtari, A. (2019).
\newblock \enquote{R-squared for {B}ayesian regression models.}
\newblock {\em The American Statistician\/}, 73(3): 307--309.
\endbibitem

\bibitem[{Goutis and Robert(1998)}]{goutis1998}
Goutis, C. and Robert, C.~P. (1998).
\newblock \enquote{{Model choice in generalised linear models: A Bayesian
  approach via Kullback-Leibler projections}.}
\newblock {\em Biometrika\/}, 85(1): 29--37.
\endbibitem

\bibitem[{Kaufman and Sain(2010)}]{kaufman2010}
Kaufman, C.~G. and Sain, S.~R. (2010).
\newblock \enquote{{B}ayesian functional {ANOVA} modeling using {G}aussian
  process prior distributions.}
\newblock {\em Bayesian Analysis\/}, 5(1): 123--149.
\endbibitem

\bibitem[{Kimeldorf and Wahba(1970)}]{kimeldorf1970}
Kimeldorf, G.~S. and Wahba, G. (1970).
\newblock \enquote{{A Correspondence Between Bayesian Estimation on Stochastic
  Processes and Smoothing by Splines}.}
\newblock {\em The Annals of Mathematical Statistics\/}, 41(2): 495 -- 502.
\endbibitem

\bibitem[{Liu et~al.(2020)Liu, Ong, Shen, and Cai}]{liu2020}
Liu, H., Ong, Y.-S., Shen, X., and Cai, J. (2020).
\newblock \enquote{When {G}aussian Process Meets Big Data: A Review of Scalable
  {GP}s.}
\newblock {\em IEEE Transactions on Neural Networks and Learning Systems\/},
  31(11): 4405--4423.
\endbibitem

\bibitem[{Lu et~al.(2022)Lu, Boukouvalas, and Hensman}]{lu22}
Lu, X., Boukouvalas, A., and Hensman, J. (2022).
\newblock \enquote{Additive {G}aussian Processes Revisited.}
\newblock In Chaudhuri, K., Jegelka, S., Song, L., Szepesvari, C., Niu, G., and
  Sabato, S. (eds.), {\em Proceedings of the 39th International Conference on
  Machine Learning\/}, volume 162 of {\em Proceedings of Machine Learning
  Research\/}, 14358--14383. PMLR.
\endbibitem

\bibitem[{McLatchie et~al.(2024)McLatchie, Rögnvaldsson, Weber, and
  Vehtari}]{mclatchie2024}
McLatchie, Y., Rögnvaldsson, S., Weber, F., and Vehtari, A. (2024).
\newblock \enquote{Advances in projection predictive inference.}
\newline\urlprefix\url{https://arxiv.org/abs/2306.15581}
\endbibitem

\bibitem[{{MIT Election Data and Science Lab}(2017)}]{mit2017}
{MIT Election Data and Science Lab} (2017).
\newblock \enquote{{U.S. President 1976–2020, V6}.}
\endbibitem

\bibitem[{Pavone et~al.(2020)Pavone, Piironen, Burkner, and
  Vehtari}]{pavone2020}
Pavone, F., Piironen, J., Burkner, P.-C., and Vehtari, A. (2020).
\newblock \enquote{Using reference models in variable selection.}
\newblock {\em Computational Statistics\/}, 38: 349--371.
\endbibitem

\bibitem[{Pedersen et~al.(2019)Pedersen, Miller, Simpson, and
  Ross}]{pedersen2019}
Pedersen, E., Miller, D., Simpson, G., and Ross, N. (2019).
\newblock \enquote{Hierarchical generalized additive models in ecology: An
  introduction with mgcv.}
\newblock {\em PeerJ\/}, (5).
\endbibitem

\bibitem[{Piironen et~al.(2020)Piironen, Paasiniemi, and
  Vehtari}]{piironen2020}
Piironen, J., Paasiniemi, M., and Vehtari, A. (2020).
\newblock \enquote{{Projective inference in high-dimensional problems:
  Prediction and feature selection}.}
\newblock {\em Electronic Journal of Statistics\/}, 14(1): 2155 -- 2197.
\endbibitem

\bibitem[{Piironen and Vehtari(2017)}]{piironen2017}
Piironen, J. and Vehtari, A. (2017).
\newblock \enquote{Comparison of Bayesian predictive methods for model
  selection.}
\newblock {\em Statistics and Computing\/}, 27(3): 711--735.
\endbibitem

\bibitem[{Pinheiro and Bates(2000)}]{pinheiro2000}
Pinheiro, J.~C. and Bates, D.~M. (2000).
\newblock {\em Mixed-effects models in {S} and {S-PLUS}\/}.
\newblock New York, NY: Springer.
\endbibitem

\bibitem[{Qian et~al.(2008)Qian, Wu, and Wu}]{qian2008}
Qian, P. Z.~G., Wu, H., and Wu, C. F.~J. (2008).
\newblock \enquote{{G}aussian Process Models for Computer Experiments With
  Qualitative and Quantitative Factors.}
\newblock {\em Technometrics\/}, 50(3): 383--396.
\endbibitem

\bibitem[{Qui\~{n}onero Candela and Rasmussen(2005)}]{quinonerocandela2005}
Qui\~{n}onero Candela, J. and Rasmussen, C.~E. (2005).
\newblock \enquote{A Unifying View of Sparse Approximate {G}aussian Process
  Regression.}
\newblock {\em J. Mach. Learn. Res.\/}, 6: 1939--1959.
\endbibitem

\bibitem[{Quintana et~al.(2016)Quintana, Johnson, Waetjen, and
  Gold}]{quintana2016}
Quintana, F.~A., Johnson, W.~O., Waetjen, L.~E., and Gold, E.~B. (2016).
\newblock \enquote{{B}ayesian Nonparametric Longitudinal Data Analysis.}
\newblock {\em Journal of the American Statistical Association\/}, 111(515):
  1168--1181.
\endbibitem

\bibitem[{{R Core Team}(2023)}]{rcore2023}
{R Core Team} (2023).
\newblock {\em R: A Language and Environment for Statistical Computing\/}.
\newblock R Foundation for Statistical Computing, Vienna, Austria.
\newline\urlprefix\url{https://www.R-project.org/}
\endbibitem

\bibitem[{Ramsay and Silverman(2005)}]{ramsay2005}
Ramsay, J. and Silverman, B.~W. (2005).
\newblock {\em Functional Data Analysis\/}.
\newblock Springer, New York, NY, 2nd edition.
\endbibitem

\bibitem[{Rasmussen and Williams(2006)}]{rasmussen2006}
Rasmussen, C.~E. and Williams, C. K.~I. (2006).
\newblock {\em {{{G}aussian} Processes for Machine Learning}\/}.
\newblock Cambridge, Massachusetts: MIT Press.
\endbibitem

\bibitem[{Riutort-Mayol et~al.(2022)Riutort-Mayol, B\"{u}rkner, Andersen,
  Solin, and Vehtari}]{riutortmayol2020}
Riutort-Mayol, G., B\"{u}rkner, P.-C., Andersen, M.~R., Solin, A., and Vehtari,
  A. (2022).
\newblock \enquote{Practical {H}ilbert space approximate {B}ayesian {G}aussian
  processes for probabilistic programming.}
\newblock {\em Statistics and Computing\/}, 33(1).
\endbibitem

\bibitem[{Roustant et~al.(2020)Roustant, Padonou, Deville, Clément, Perrin,
  Giorla, and Wynn}]{roustant2018}
Roustant, O., Padonou, E., Deville, Y., Clément, A., Perrin, G., Giorla, J.,
  and Wynn, H. (2020).
\newblock \enquote{Group Kernels for {G}aussian Process Metamodels with
  Categorical Inputs.}
\newblock {\em SIAM/ASA Journal on Uncertainty Quantification\/}, 8(2):
  775--806.
\endbibitem

\bibitem[{Sacks et~al.(1989)Sacks, Welch, Mitchell, and Wynn}]{sacks1989}
Sacks, J., Welch, W.~J., Mitchell, T.~J., and Wynn, H.~P. (1989).
\newblock \enquote{Design and Analysis of Computer Experiments.}
\newblock {\em Statistical Science\/}, 4(4): 409--423.
\endbibitem

\bibitem[{Shmueli(2010)}]{shmueli2010}
Shmueli, G. (2010).
\newblock \enquote{{To Explain or to Predict?}}
\newblock {\em Statistical Science\/}, 25(3): 289 -- 310.
\endbibitem

\bibitem[{Snelson and Ghahramani(2006)}]{snelson2006}
Snelson, E. and Ghahramani, Z. (2006).
\newblock \enquote{Sparse {G}aussian Processes using Pseudo-inputs.}
\newblock In {\em Advances in Neural Information Processing Systems\/},
  volume~18.
\endbibitem

\bibitem[{Snoek et~al.(2014)Snoek, Swersky, Zemel, and Adams}]{snoek2014}
Snoek, J., Swersky, K., Zemel, R., and Adams, R.~P. (2014).
\newblock \enquote{Input warping for Bayesian optimization of non-stationary
  functions.}
\newblock In {\em Proceedings of the 31st International Conference on
  International Conference on Machine Learning\/}, volume~32.
\endbibitem

\bibitem[{Solin and S{\"a}rkk{\"a}(2020)}]{solin2019}
Solin, A. and S{\"a}rkk{\"a}, S. (2020).
\newblock \enquote{Hilbert space methods for reduced-rank {G}aussian process
  regression.}
\newblock {\em Statistics and Computing\/}, 30: 419--446.
\endbibitem

\bibitem[{Timonen et~al.(2021)Timonen, Mannerström, Vehtari, and
  Lähdesmäki}]{timonen2021}
Timonen, J., Mannerström, H., Vehtari, A., and Lähdesmäki, H. (2021).
\newblock \enquote{{lgpr: an interpretable non-parametric method for inferring
  covariate effects from longitudinal data}.}
\newblock {\em Bioinformatics\/}, 37(13): 1860--1867.
\endbibitem

\bibitem[{Titsias(2009)}]{titsias2009}
Titsias, M. (2009).
\newblock \enquote{Variational Learning of Inducing Variables in Sparse
  {G}aussian Processes.}
\newblock In {\em Proceedings of the 12th International Conference on
  Artificial Intelligence and Statistics\/}, volume~5.
\endbibitem

\bibitem[{Trangucci(2017)}]{trangucci2017}
Trangucci, R. (2017).
\newblock \enquote{Hierarchical {G}aussian Processes in {S}tan.}
\newline\urlprefix\url{https://mc-stan.org/events/stancon2017-notebooks/stancon2017-trangucci-hierarchical-gps.pdf}
\endbibitem

\bibitem[{Vehtari et~al.(2017)Vehtari, Gelman, and Gabry}]{vehtari2017}
Vehtari, A., Gelman, A., and Gabry, J. (2017).
\newblock \enquote{Practical {B}ayesian model evaluation using leave-one-out
  cross-validation and {WAIC}.}
\newblock {\em Statistics and Computing\/}, 27(5): 1413--1432.
\endbibitem

\bibitem[{Verbeke and Molenberghs(2000)}]{verbeke2000}
Verbeke, G. and Molenberghs, G. (2000).
\newblock {\em Linear Mixed Models for Longitudinal Data\/}.
\newblock Springer, New York, NY.
\endbibitem

\bibitem[{Wang et~al.(2021)Wang, Yerramilli, Iyer, Apley, Zhu, and
  Chen}]{wang2021}
Wang, L., Yerramilli, S., Iyer, A., Apley, D., Zhu, P., and Chen, W. (2021).
\newblock \enquote{{Scalable Gaussian Processes for Data-Driven Design Using
  Big Data With Categorical Factors}.}
\newblock {\em Journal of Mechanical Design\/}, 144(2).
\endbibitem

\bibitem[{Wilson and Nickisch(2015)}]{wilson15}
Wilson, A.~G. and Nickisch, H. (2015).
\newblock \enquote{Kernel interpolation for scalable structured Gaussian
  processes (KISS-GP).}
\newblock In {\em Proceedings of the 32nd International Conference on
  International Conference on Machine Learning\/}, volume~37.
\endbibitem

\bibitem[{Wood(2017)}]{wood2017}
Wood, S. (2017).
\newblock {\em Generalized Additive Models: {A}n Introduction with {R}\/}.
\newblock Chapman and Hall/CRC, 2nd edition.
\endbibitem

\bibitem[{Zhang and Notz(2015)}]{zhang2015}
Zhang, Y. and Notz, W.~I. (2015).
\newblock \enquote{Computer Experiments with Qualitative and Quantitative
  Variables: A Review and Reexamination.}
\newblock {\em Quality Engineering\/}, 27: 2--13.
\endbibitem

\bibitem[{Zhang et~al.(2020)Zhang, Tao, Chen, and Apley}]{zhang2020}
Zhang, Y., Tao, S., Chen, W., and Apley, D.~W. (2020).
\newblock \enquote{A Latent Variable Approach to {G}aussian Process Modeling
  with Qualitative and Quantitative Factors.}
\newblock {\em Technometrics\/}, 62(3): 291--302.
\endbibitem

\end{thebibliography}

\includepdf[pages=-]{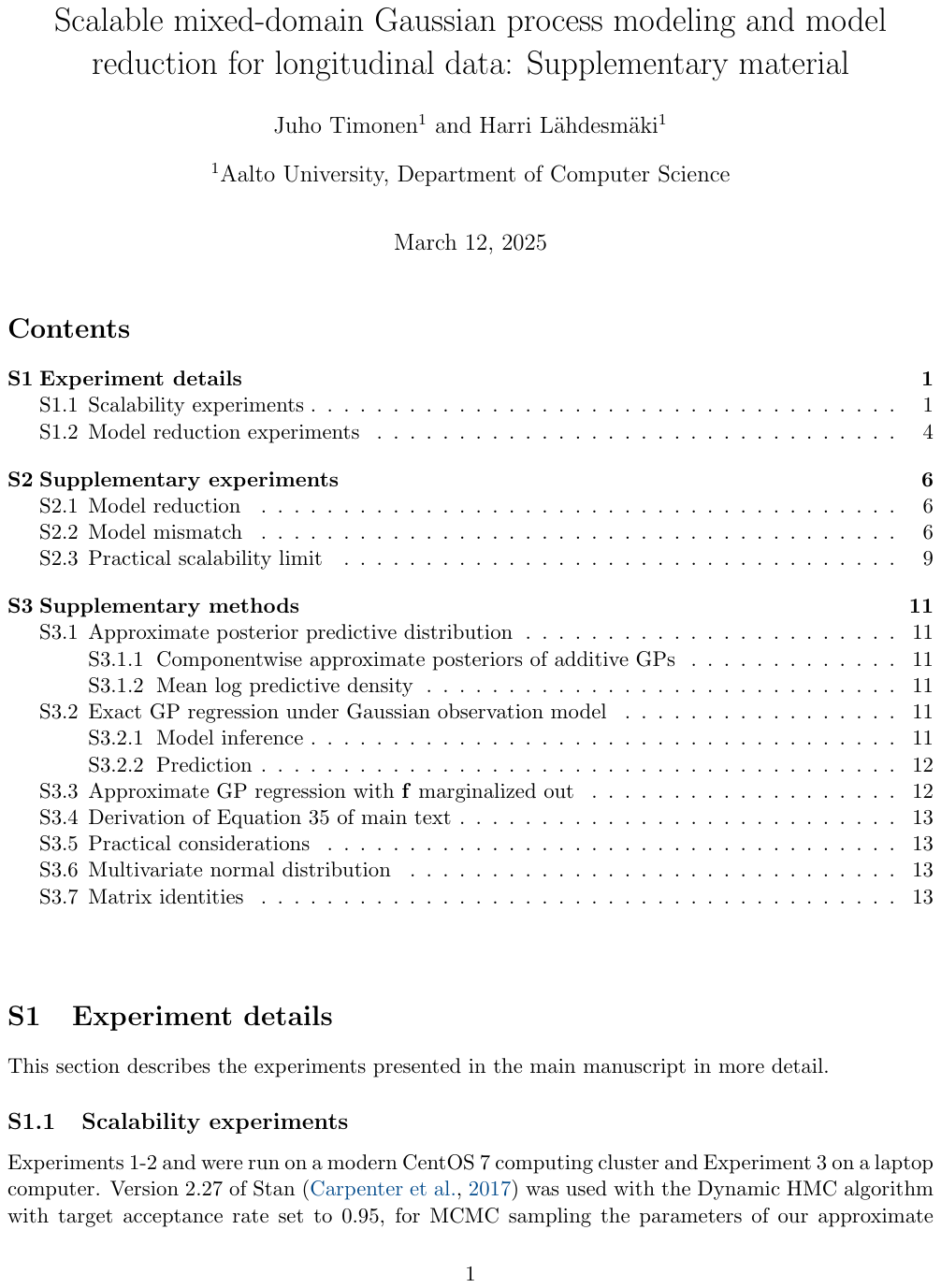}

\end{document}